\documentclass{iucr}
\usepackage{amssymb}
\usepackage{amsmath}
\usepackage{graphicx}

\paperprodcode{a000000}
  \paperref{xx9999}
  \papertype{FA}
  \paperlang{english}
  \journalcode{S}
  \journalyr{2012}
  \journaliss{6}
  \journalvol{19}
  \journalfirstpage{001}
  \journallastpage{007}
  \journalreceived{0 XXXXXXX 2012}
  \journalaccepted{0 XXXXXXX 2012}
  \journalonline{0 XXXXXXX 2012}

\begin{document}

\title{Spectrometer for Hard X-Ray Free Electron Laser Based on Diffraction Focusing}
\shorttitle{Spectrometer for Hard XFEL}
\author[a]{V. G.}{Kohn}{}{}
\author[b]{O. Y.}{Gorobtsov}{}{}
\cauthor[b,c]{I. A.}{Vartanyants}{ivan.vartaniants@desy.de}{}
\aff[a]{National Research Center "Kurchatov Institute", Kurchatov square 1, 123182 \city{Moscow} \country{Russia} }
\aff[b]{Deutsches Electronen-Synchrotron DESY, Notkestr. 85, D-22607 \city{Hamburg} \country{Germany} }
\aff[c]{National Research Nuclear University, "MEPhI", 115409 \city{Moscow} \country{Russia} }
\maketitle

\date{today}

\begin{synopsis}
We propose the diffraction focusing spectrometer (DFS) which is able to
measure the whole energy spectrum of the radiation at the same time.
It is well suited for measurements of the single pulses of hard x-ray
FELs.
\end{synopsis}

\begin{abstract}
X-ray free electron lasers (XFELs) generate sequences of ultra-short, spatially coherent pulses of x-ray radiation.
We propose the diffraction focusing spectrometer (DFS), which is able to
measure the whole energy spectrum of the radiation of a single XFEL pulse with an energy
 resolution of $\Delta E/E\approx 2\times 10^{-6}$.
This is much better than for most modern x-ray spectrometers. Such
resolution allows one to resolve the fine spectral structure of the XFEL pulse. The effect
of diffraction focusing occurs in a single crystal plate due to dynamical scattering,
and is similar to focusing in a Pendry lens made from the metamaterial with a
negative refraction index. Such a spectrometer is easier to operate
than those based on bent crystals. We show that the DFS can be used in a wide
energy range from 5 keV to 20 keV.
\end{abstract}

\section{Introduction}

Hard x-ray free electron lasers (XFELs) (Emma et al. (2010), Ishikawa et al. (2012), Altarelli et al. (2006)) generate sequences of ultra-short, spatially coherent pulses of x-ray radiation. These pulses can be less than 100 fs in duration, and have an energy spectrum of the relative width $\Delta\omega/\omega\sim 10^{-3}$ ( see for example Altarelli et al. (2006)). Such characteristics make XFELs useful for a wide range of applications, including potentially 3D imaging of individual macromolecules, looking at dynamics of chemical reactions and many others. It is well known, that the XFEL pulse energy spectrum has a fine structure, which changes chaotically from pulse to pulse. This fine structure has a characteristic width on the order of $\Delta\omega/\omega\sim 10^{-6}$ for hard x-ray energies. Therefore, it is of interest to elaborate a spectrometer with sufficient resolution to resolve these fine features for each pulse individually.

Since each pulse of the XFEL has a unique structure of the energy spectrum, the spectrometer has to be able to detect the whole spectrum of a single pulse simultaneously. Obviously, the scan of the energy region by the monochromator, as it is conventionally performed at synchrotron sources, can not be applied at XFELs. Most of the known spectrometers (for example, Kleimenov et. al. (2009), Dickinson et. al. (2008)) of such type have the maximum energy resolution of $10^{-5}$. They are based on the Bragg diffraction of the curved or flat single crystals, and their resolution is limited by the angular divergence of the reflected beam due to the dynamical scattering effects. Recently, a spectrometer based on a curved crystal in the Bragg scattering geometry for applications at XFEL sources was developed (Zhu, 2012). It still has a resolving power of $2.4\times10^{-5}$.

We propose a new spectrometer based on the effect of a single crystal diffraction focusing (Afanasiev \& Kohn, 1977). A sketch of the diffraction focusing spectrometer (DFS) is shown in Fig. 1. In this spectrometer the incoming divergent beam from a secondary source is focused by a single crystal plate at each energy  near the Bragg angle. Since the Bragg angle depends on the energy, x-rays with different energy will be focused at different points just behind the crystal under conditions of focusing. Due to this property a higher resolution  spectrometer can be developed.
\begin{figure}
\centering
\includegraphics{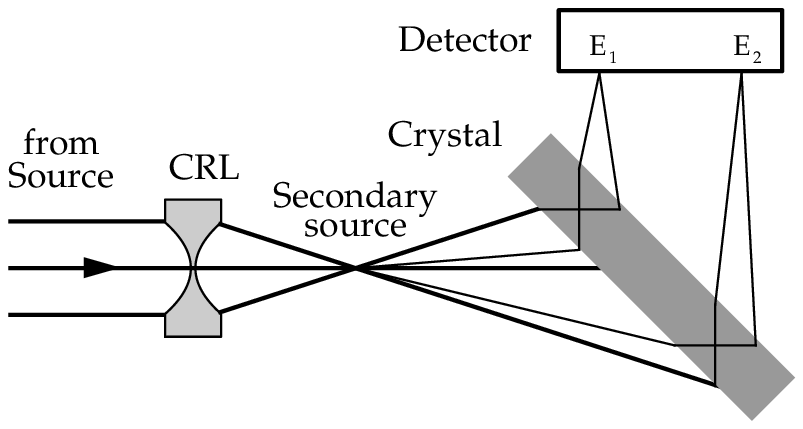}
\caption{Schematic view of the beamline with the diffraction focusing spectrometer. The incoming beam from the source is focused by the compound refractive lenses (CRLs) and  creates the secondary source. The flat crystal is positioned in a Laue geometry behind this secondary source. The high resolution detector is located close to the crystal in the direction of the diffracted beam.}%
\label{fig1}
\end{figure}
In order to achieve high resolution, the effective size of the secondary source should be decreased to about few microns and the angular divergence of the beam increased to the necessary value. This can be done, for example, by positioning compound refractive lenses (CRLs) (Snigirev et al., 1996; 1998) upstream from the DFS. Planar Silicon CRLs allow one to achieve small focusing distance (few centimeters), and obtain the source image of the small size even at high photon energies.

An effect of single crystal diffraction focusing was first predicted by Afanasiev \& Kohn (1977) and considered in detail by Kohn (1979). It was experimentally confirmed by Aristov et al. (1978, 1980, 1982), and by Koz'mik \& Mikhailyuk (1978). Recently the effect of focusing in bent-crystal systems was considered theoretically by Mocella et al. (2004, 2008), and by Nesterets \& Wilkins (2008) as a way for improving polychromatic focusing.

The effect of diffraction focusing is similar to one predicted in metamaterials with a negative refractive index for a long wavelength radiation. For these materials Pendry (2000) has introduced the lens in the form of a plane slab made from the metamaterial. The divergent incident beam is focused, first inside the slab, and then behind it.

For x-rays an effective negative refraction occurs for a wave, for which, due to dynamical diffraction in the Laue case, the Pointing vector has an opposite direction to the angular deviation of the incident wave from the Bragg angle.
As a result, a slightly divergent beam incident at the Bragg angle at the single crystal is focused, first inside the crystal, and then once again behind the crystal. The difference from the Pendry lens is that the angular width of the focused beam is equal to the angular range of the dynamical diffraction (tens of microradians), and is rather small. Nevertheless, this effect allows one to obtain a beam of few microns width. The width of the focused beam determines the resolution of the DFS. In this work we investigate the optimal conditions for such a spectrometer based on diffraction focusing and show that the energy resolution of DFS can reach $10^{-6}$. Such resolution will be sufficient to resolve the fine structure of the hard x-ray FEL pulse spectrum.
\\
\begin{figure}
\centering
\includegraphics{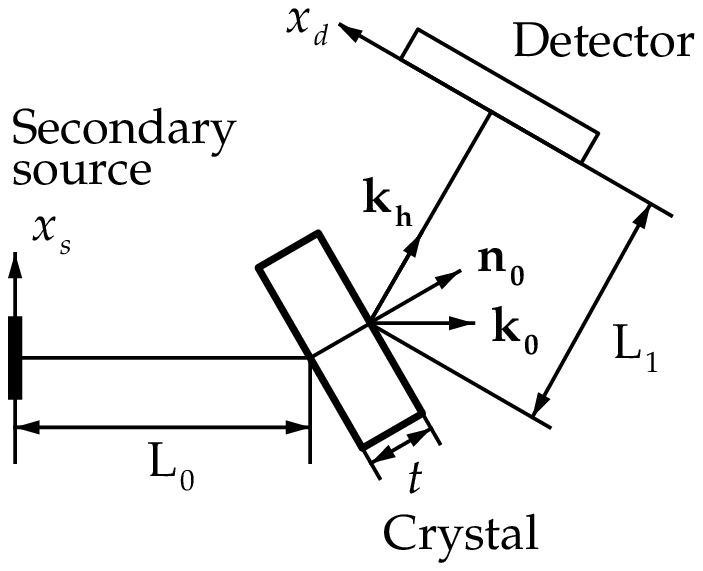}
\caption{Main components of the diffraction focusing spectrometer. The secondary source plane is described by the coordinate $x_s$, crystal is positioned at the distance $L_0$ from the source, detector is located at the distance $L_1$ from the crystal in the direction of the Laue diffracted beam defined by the vector $\mathbf{k_h}$. The detector plane is described by the coordinate $x_d$.}%
\label{fig2}
\end{figure}
\hspace*{\fill}

\section{Theory and Analytical Estimates of the Achievable Resolution}

Our goal is to obtain an estimate of the resolution of the spectrometer based on diffraction focusing.
To get this value it is
necessary to calculate the full width at half maximum (FWHM) of the
intensity profile of the focused monochromatic beam at the detector position.
Main components of the DFS are shown in Fig.2. The crystal in the
form of plane parallel plate of thickness $t$ is located at the distance $%
L_{0}$ from the point source located at the coordinate $x_{s}$ in the plane
of  the secondary source formed by the CRL (Fig. 1).
The detector plane is positioned at the distance $L_{1}$ and is perpendicular to the direction
of the diffracted beam with the wave vector $\mathbf{k_{h}}$. The coordinate $%
x_{d}$ is related to the point of observation at the detector. The angles $%
\varphi _{0}$ and $\varphi _{h}$ between the normal to the crystal surface $%
\mathbf{n_{0}}$ and the wave vectors $\mathbf{k_{0}}$, $\mathbf{k_{h}}$
determine the asymmetry of the diffraction; their sum is equal to
$2\theta _{B}$, where $\theta _{B}$\ is the Bragg angle.

The amplitude of the diffracted wave $E_{h}(x_d)$ at the
detector position in the scheme shown in Fig. 2 was obtained by Afanasev \& Kohn (1971). It was derived by introducing the crystal propagator in the form of the Green function for the Takagi equations (Authier, 2005), and the Kirchhoff
propagator, describing the transmission in free space in paraxial approximation. In the following paper by Kohn et al. (2000) the final form of this amplitude was obtained by using Fourier transform in reciprocal space. Using this approach the amplitude of the diffracted wave $E_{h}(x_d)$ can be presented in the following way
\begin{equation}
E_{h}(x_d) = C_{1}\int\limits_{-\infty}^{\infty} \frac{d\eta }{S(\eta )}\sum_{\pm }\left[ \pm \exp (i\Phi
_{\pm }(\eta )-M_{\pm }(\eta ))\right] .
\label{01}
\end{equation}
Here the positive sign corresponds to the weakly absorbing wave, and the negative sign
 to the strongly absorbing wave,
\begin{eqnarray}
\Phi _{\pm }(\eta ) &=&\frac{\pi \alpha }{\Lambda }\left( x_{d}-x_{C}-\frac{%
x_{s}}{\beta }\right) \eta -\frac{\pi t_{0}}{2\Lambda }\eta ^{2}+\Phi _{\pm
}^{(0)}(\eta ),  \notag \\
\Phi _{\pm }^{(0)}(\eta ) &=&\pm \frac{\pi t}{\Lambda }S(\eta ),\quad S(\eta
)=(1+\eta ^{2})^{1/2},  \label{02} \\
M_{\pm }(\eta ) &=&\frac{\mu _{0}t}{2(\gamma _{0}\gamma _{h})^{1/2}}\left[
\frac{(\beta +1)}{2\beta ^{1/2}}\mp \frac{\varepsilon _{h}}{S(\eta )}\mp
\frac{\eta }{S(\eta )}\frac{(\beta -1)}{2\beta ^{1/2}}\right] ,  \notag \\
C_{1} &=&\frac{(\lambda L_{t})^{1/2}}{2S_{B}\Lambda \beta ^{1/2}},\quad \mu
_{0}=K\chi _{i0},\quad K=\frac{2\pi }{\lambda }.  \notag
\end{eqnarray}%
In these equations the following parameters are used: $\lambda $ is the wavelength, $\mu _{0}$ is the absorption coefficient,
$L_{t}=L_{0}+L_{1}$, $S_{B}=\sin (2\theta _{B})$,
\begin{eqnarray}
\Lambda &=&\frac{\lambda (\gamma _{0}\gamma _{h})^{1/2}}{|\chi _{rh}|},\quad
t_{0}=\frac{2|\chi _{rh}|\gamma _{0}^{3/2}}{S_{B}^{\,2}\gamma _{h}^{1/2}}%
\widetilde{L},\quad \widetilde{L}=\frac{L_{0}}{\beta ^{2}}+L_{1},  \notag \\
x_{C} &=&x_{0}-x_{L},\quad x_{L}=\frac{\lambda \widetilde{L}q_{0}}{2\pi }%
,\quad x_{0}=\frac{1}{2}t\gamma _{h}(T_{h}-T_{0}),  \label{03} \\
q_{0} &=&\frac{K|\chi _{r0}|}{2S_{B}}(\beta -1),\quad \varepsilon _{h}=\frac{%
\chi _{ih}}{\chi _{i0}},\quad \alpha =\frac{2\gamma _{0}}{S_{B}},  \notag \\
T_{0,h} &=&\tan (\varphi _{0,h}),\quad \gamma _{0,h}=\cos (\varphi
_{0,h}),\quad \beta =\frac{\gamma _{0}}{\gamma _{h}},  \notag
\end{eqnarray}

In Eqs. (2-3) $\Lambda $ is the extinction length, $t_{0}$ is an effective focusing thickness of the crystal, $\widetilde{L}$ is an effective distance from the
source to the detector,  $x_{0}$ and $x_{L}$ are the displacements of the image due to
asymmetry of the diffraction geometry,  $\chi _{0,h}=\chi _{r0,h}+i\chi _{i0,h}$ are the Fourier
components of the complex susceptibility of the crystal for the angles of
scattering $0$ and $2\theta _{B}$. The intensity distribution at the detector position $x_d$ for a point source at the position $x_s$ is given by the square modulus of the amplitude $E_h(x_d)$ (Eq. (1)), $I_h(x_d)=|E_h(x_d)|^2$. For geometrical optics description of the effect of diffraction focusing see Appendix.

The DFS energy resolution is determined by the condition that a shift of the photon
energy $\Delta E$ leads to a shift of the diffracted beam center by $x_{d\omega }$ at
the detector plane. It was shown by Kohn et al. (2000) that%
\begin{equation}
x_{d\omega }=(\frac{L_{0}}{\beta }-L_{1})\tan \theta _{B}\frac{\Delta E}{E}.
\label{04}
\end{equation}
We will consider now the case of the crystal thickness $t=t_0$, when focusing conditions are satisfied at the detector position. To obtain an estimate of the DFS resolution the shift of the beam position $x_{d\omega }$ in equation (\ref{04}) has to be compared with the beam size (FWHM) $\Delta _{x}$ at the detector position. To obtain an analytical estimate of the beam size we will consider two cases of weakly absorbing ($\mu _{0}t_{0} \ll 1$) and strongly absorbing ($\mu _{0}t_{0} \gg 1$) crystals.

In the first case, we can neglect absorption and consider in equation (\ref{01}) the central part
of the integration region with $\eta \ll 1$. This approximation is justified if the additional condition $\Lambda \ll t_{0}$ is satisfied as well. Then, we can consider only one term in the equation (\ref{01}) with
the positive sign, expand $S(\eta )$ in the Taylor series up to the fourth order in $\eta$ and
obtain
\begin{equation}
E_{h}(x_d) \approx C_{1}\int\limits_{-\infty}^{\infty} d\eta \exp \left( i\pi \eta s+i\frac{\pi (t-t_{0})}{%
\Lambda }\frac{\eta ^{2}}{2}-i\frac{\pi t}{\Lambda }\frac{\eta ^{4}}{8}%
\right) ,  \label{05}
\end{equation}
where
\begin{equation}
s=\frac{\alpha }{\Lambda }\left( x_{d}-x_{C}-\frac{x_{s}}{\beta }\right)
\label{06}
\end{equation}
is the dimensionless coordinate at the detector position. As soon as we consider small values of $\eta$, the
integral in the equation (\ref{05}) reaches its maximum value if the following conditions $t=t_{0}$, and $s=0$ are satisfied simultaneously. In fact, the first condition is the condition of focusing. It also defines the
focusing crystal thickness $t_{0}$, that is according to its definition in equation (\ref{03}), is proportional to the effective distance from the source to the detector $\widetilde{L}$. This effective distance is equal to the total
distance $L_{t}$\ in the symmetrical ($\beta =1$) scattering geometry.

The integral in equation (\ref{05}) is similar to the one describing the diffraction by a
parabolic mirror (Berry \& Klein, 1996). Therefore, the diffraction pattern
in the case of small absorption, when the phase term plays the main role,
is similar to the one resulting from the parabolic mirror focusing.

To obtain an estimate of the focused beam size, we note that the exponential in the
integrand becomes a fast oscillating function for small values of $s$
outside the region where the term $\eta ^{4}$ is less than $\pi $. Therefore,
we can exclude this region from the integration, and consider the integral
in the finite limits from $-\eta _{0}$ to $\eta _{0}$, where
\begin{equation}
\eta _{0}\approx (8\Lambda /t_{0})^{1/4} . \label{07}
\end{equation}
Then we have for the amplitude in equation (\ref{05})
\begin{equation}
\frac{E_{h} (x_d)}{C_{1}} \approx \int\limits_{-\eta _{0}}^{\eta _{0}}d\eta \exp \left(
i\pi \eta s\right) =2\eta _{0}\frac{\sin (\pi \eta _{0}s)}{\pi \eta _{0}s}.
\label{08}
\end{equation}
Thus, we arrive to the conclusion that the amplitude of the focused wave in a weak absorption case is described
by the sinc-function, for which the size (FWHM) of the intensity distribution  is given by
\begin{equation}
\Delta _{xp}=0.83\frac{\Lambda ^{3/4}t_{0}^{1/4}}{\pi \alpha }.
\label{09}
\end{equation}
\begin{figure}
\centering
\includegraphics{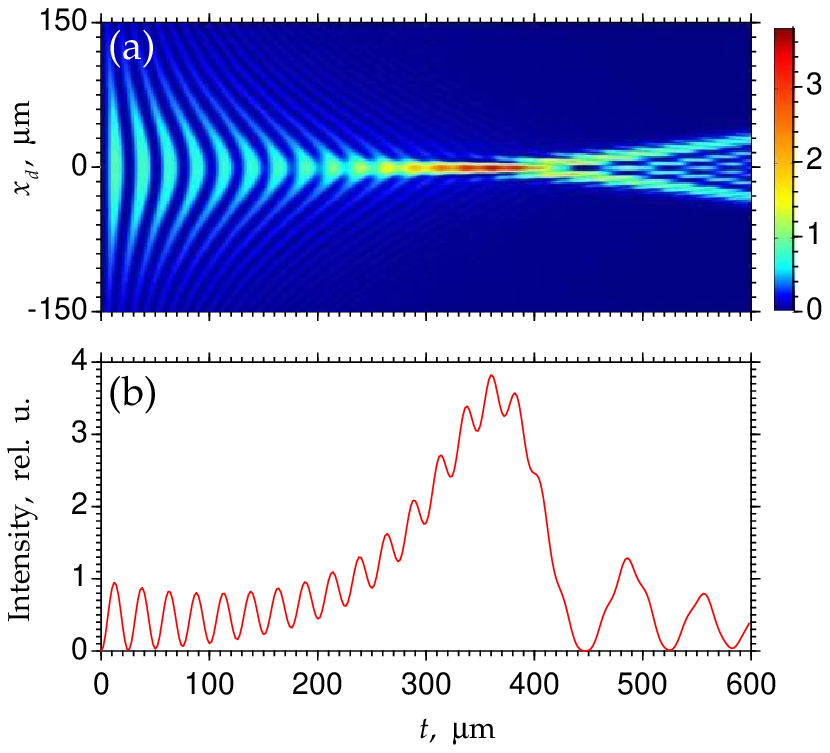}
\caption{The case of a Silicon 220 reflection. (a) Two-dimensional distribution of the intensity. (b) Intensity in the central position of the image $x_d=0$ as a function of the crystal thickness $t$.}
\label{fig3}
\end{figure}
The second case of strong absorption was discussed earlier by Afanasev \& Kohn
(1977). Again, in equation (\ref{01}) we take into consideration only one term with the positive sign for which the
Borrmann effect takes place in the region of small $\eta \ll 1$. In this
case the absorption term can not be neglected, and we can expand functions $\Phi _{+}(\eta )$ and $%
M_{+}(\eta )$\ in the exponent of the Taylor series up to the second order in $\eta$ and obtain for the amplitude
\begin{equation}
E_{h}(x_d) \approx C_{2}{\int\limits_{-\infty}^{\infty} }d\eta \exp (i\pi s\eta -\sigma _{1}\eta ^{2}+\sigma
_{2}\eta ),  \label{10}
\end{equation}%
where%
\begin{eqnarray}
C_{2} &=&C_{1}\exp \left( -\frac{\mu _{0}t_{0}}{2(\gamma _{0}\gamma
_{h})^{1/2}}\left[ \frac{(\beta +1)}{2\beta ^{1/2}}-\varepsilon _{h}\right]
\right) ,  \notag \\
\sigma _{1} &=&\frac{\mu _{0}t_{0}\varepsilon _{h}}{4(\gamma _{0}\gamma
_{h})^{1/2}},\quad \sigma _{2}=\frac{(\beta -1)\mu _{0}t_{0}}{4\gamma _{0}}.
\label{11}
\end{eqnarray}%

The integral (\ref{10}) is similar to the well known Fourier transform of the
Kirchhoff propagator%
\begin{equation}
\int\limits_{-\infty}^{\infty} \frac{dq}{2\pi }\exp \left( iqx-i\frac{\lambda z}{4\pi }q^{2}\right) =%
\frac{1}{(i\lambda z)^{1/2}}\exp \left( i\pi \frac{x^{2}}{\lambda z}\right)
\label{12}
\end{equation}%
and can be calculated analytically as%
\begin{eqnarray}
E_{h}(x_d) &\approx&C_{3}\exp \left( i2\pi \sigma _{2}s-\frac{\pi ^{2}}{4\sigma _{1}}%
s^{2}\right) ,  \notag \\
C_{3} &=&C_{2}\left( \frac{\pi }{\sigma _{1}}\right) ^{1/2}\exp \left( -%
\frac{\sigma _{2}^{2}}{4\sigma _{1}}\right).
\label{13}
\end{eqnarray}%
Now the size (FWHM) of the intensity distribution in the focus for a strong absorption case is equal to%
\begin{equation}
\Delta _{xa}=(2\ln 2)^{1/2}\frac{\Lambda }{\pi \alpha }\sigma _{1}^{1/2}.
\label{14}
\end{equation}

Direct computer simulations show that both values of the peak size $\Delta _{xp}$ and $\Delta
_{xa}$, defined by the equations (\ref{09}) and (\ref{14}), give an excellent estimate within the region of their validity. However, in some intermediate cases the region of integration is limited by the
phase but absorption can not be neglected inside this region. In these cases there is no analytical solution for the amplitude of the diffracted wave and for the estimate of the beam size one should calculate the following integral numerically
\begin{equation}
E_{h}(x_d) = C_{2}{\int\limits_{-\eta _{0}}^{\eta _{0}}}d\eta \exp (i\pi s\eta
-\sigma _{1}\eta ^{2}+\sigma _{2}\eta ),  \label{15}
\end{equation}%
and then estimate $\Delta _{x}$ as the FWHM of the intensity curve.

Finally, the energy resolution of DFS can be obtained using the Rayleigh criterium (Born \& Wolf, 2000), i.e.
two peaks are resolved, if the intermediate point between them has a
height less than 80\% of the maximum height. Taking into account equation (\ref{04}) we obtain for the energy resolution
\begin{equation}
\frac{\Delta E}{E}=\frac{1.15\beta \Delta _{x}}{(L_{0}-L_{1}\beta )\tan \theta _{B}} .
\label{16}
\end{equation}
To increase resolution one has to consider $L_{0}$ to be as large as
possible, whereas $L_{1}$ considered to be as small as possible.
Analytical results obtained for the beam size $\Delta _{x}$ (equations (\ref{09}) and (\ref{14})) allow one to estimate the
influence of the asymmetry factor $\beta$.
\begin{figure}
\centering
\includegraphics{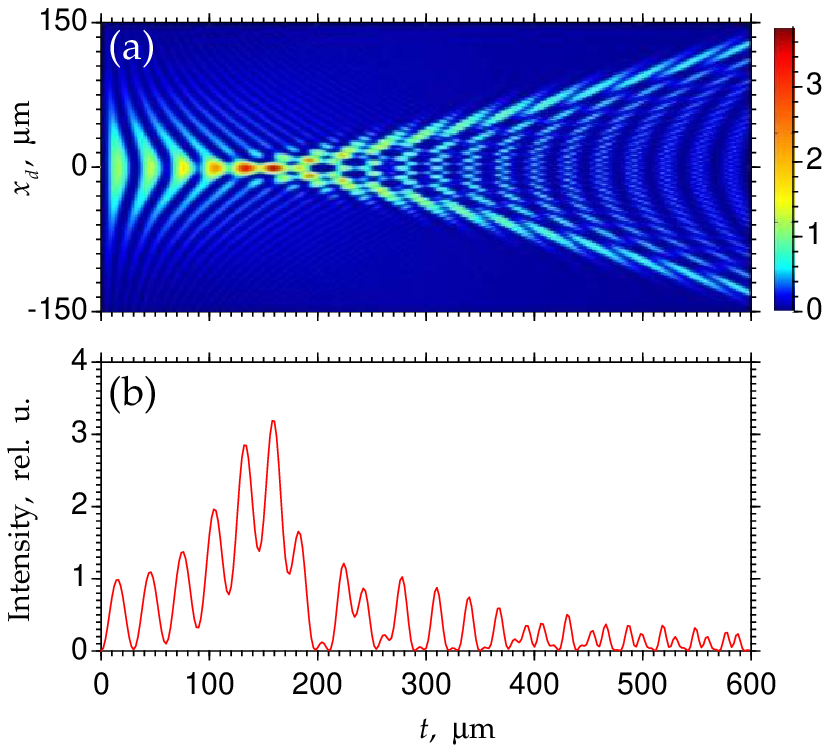}
\caption{The case of a Diamond 220 reflection (a) Two-dimensional intensity distribution. (b) Intensity in the central position at $x_d=0$ as a function of the crystal thickness $t$.}
\label{fig4}
\end{figure}

In the case of low absorption substituting equation (\ref{03})
into equation (\ref{09}) and considering the distance $L_{1}=0$ we obtain for the beam size (FWHM)
\begin{equation}
\Delta _{xp}=\frac{0.83}{2\pi }\frac{\lambda ^{3/4}S_{B}^{1/2}(2L_{0})^{1/4}%
}{|\chi _{rh}|^{1/2}\beta ^{3/4}}  .
\label{17}
\end{equation}
Substituting this equation in the expression (\ref{16}) we obtain that the resolution can be, in principle, increased in the asymmetric case with $\beta \ll 1$. Unfortunately, the gain is very small since $\Delta E/E\propto \beta ^{1/4}$.
It is also known that small values of $\beta$ are difficult to realize in practice because the surface of the crystal must be cut at the definite angle with the reflecting atomic planes close to the Bragg angle, which also depends on the radiation energy. On the other hand, as it follows from Eqs. (\ref{02}) for the small values of $\beta$ the source of the size $S$ will increase the size of the focused beam to the value $S/\beta$. This will reduce an effective resolving power of the DFS.
In the following, we will perform computer simulations and present results only for the symmetrical ($\beta =1$) scattering geometry.

\section{Results and Discussion}

In practical applications of diffraction focusing one should have
single crystals of sufficient thickness. Since large single crystals of the highest quality are
obtained from Si and Ge we considered these two cases in our simulations. Taking into account the high power of the XFEL
pulses and a demand of high thermal conductivity, we also considered a single Diamond
crystal, below referenced as C, in our simulations. Only strong
220 and 400 reflections were considered. The detailed computer simulations were
performed for the photon energy $E=12.4$ keV ($\lambda =1$ \AA). At this photon energy, the absorption coefficients are: $\mu _{0}=39.8$ cm%
$^{-1}$ for Si, $\mu _{0}=810.0$ cm$^{-1}$ for Ge, and $\mu _{0}=3.15$ cm$%
^{-1}$ for C. We note that at this photon energy and with crystal thicknesses defined by the parameter $t_0$ in equation (\ref{03}) Si
and Diamond crystals can be considered as thin ones, whereas Ge crystal is of the intermediate thickness.
\begin{figure}
\centering \includegraphics{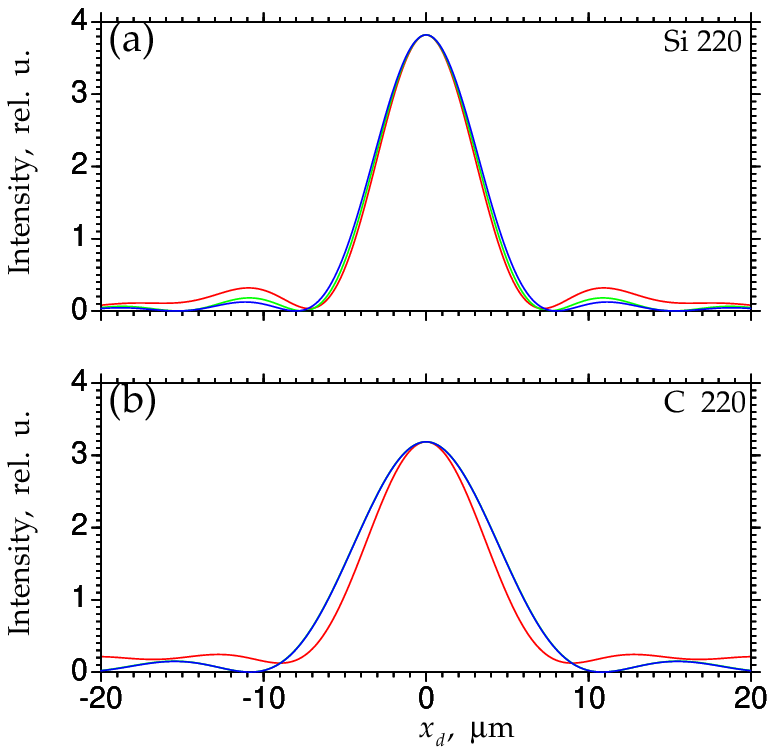}
\caption{Comparison of the intensity profiles at the detector position. (a) Silicon 220 reflection, crystal thickness $t_0 = 360$ $\mu$m. Direct calculations using Eq. (\ref{01}) (red line) and analytical results based on Eqs. (\ref{08}) (blue line) and (\ref{15}) (green line) are presented. (b) Diamond 220 reflection, crystal thickness $t_0 = 158$ $\mu$m. Direct calculation using Eq. (\ref{01}) (red line) and analytical results based on Eqs. (\ref{08}) and (\ref{15}) (blue line) are presented. Both intensity profiles obtained from the analytical result completely coincide with each other in this case. }
\label{fig5}
\end{figure}

It follows from equation (\ref{16}) that for a best DFS resolution, the distance $%
L_{0}$ should be as large as possible while the distance $L_{1}$, on the
contrary, should be as small as possible. In our simulations we considered the following distances $L_0=10$ m and $L_{1}=0.1$ m. Our simulations show that small variations of these values does not significantly influence results.

The intensity distribution at the position of the
detector as a function of the crystal thickness $t$ for the Laue diffracted beam, Si 220 reflection, and the point
source is presented in Fig. 3. The two-dimensional (2D) intensity distribution in the plane of parameters $(x_d, t)$ is presented in Fig. 3(a) and a cut through this distribution at the center of the detector position $x_d=0$ in Fig. 3(b).
Intensity values are normalized to the values at the same position along the optical axis but without a crystal.
It is well seen in this Figure that at the crystal thickness $t=t_0=360$ $\mu$m the incoming x-ray beam is focused on the exit surface of the crystal. The relative intensity is also increased to about three times comparing to a beam without a crystal. What is also well seen in Figure 3(a) are oscillations of intensity as a function of crystal thickness. They were observed before in experiments on the wedge shaped crystals (Aristov et al., 1978; 1980). For the small thicknesses $t \leq t_0$, when absorption can be neglected, these oscillations are due to a strong interference between
two types of Bloch waves (corresponding to two signs in equation (\ref{01})) inside a crystal.
The form of these oscillations is different from the ones observed with the point source at the entrance
surface of the crystal and described by Kato (1961), this is why they are called anomalous Pendell\"{o}sung effect.

The period of oscillations at $x_d=0$ (see Figure 3(a)) is equal to the
extinction length $\Lambda $ (see eq.(\ref{03})). In our scattering
conditions $\Lambda =25.1$ $\mu $m. On the other hand, the thickness of diffraction
focusing $t_{0}=360$ $\mu $m, so the condition $\Lambda \ll t_{0}$ is well satisfied.
With increasing crystal thickness the weakly absorbing field becomes larger due to
focusing effect whereas the strongly absorbing field becomes smaller due to both
defocusing and absorption effects. However, for the focusing crystal
thickness the interference between two fields still exists, and it leads to oscillations of
the peak intensity. For the conditions of the optimal resolution the crystal thickness in the focal region should be considered that provides the minimum beam size.

The results of a similar computer simulation for the case of Diamond
220 reflection are presented in Fig. 4. In this case the crystal thickness at which focusing occurs $t_{0}=158$ $\mu $m is smaller than in a Si crystal, while the extinction length is slightly larger $\Lambda =30$ $\mu $m.
The absorption is also smaller for a Diamond crystal. As a result,
strong interference fringes do not vanish at the crystal thickness corresponding to focusing conditions. Still, the
peak with the smallest width can also be observed in this case.
It is interesting that in this case for the large crystal thicknesses well pronounced Kato oscillations can be observed. They have the opposite sign of curvature comparing to the anomalous Pendell\"{o}sung fringes discussed before. One can call fringes at small crystal thicknesses as plane wave interference, and fringes at the large crystal thicknesses as
spherical wave interference.
The latter occurs at crystal thicknesses just behind the crystal thickness $t_0$ corresponding to focusing conditions, where
the focus itself can be considered as a secondary source. It is similar to
the case considered by Kato with the source at the entrance surface of the
crystal, if we separate the crystal into two parts. The first part creates the
secondary source and the second part gives the spherical wave interference.
Our simulations show that the Diamond crystal can be considered as a good candidate for a DFS because, despite the
interference effects, one can observe the focused beam with the smaller size.

We also performed simulations for Si 400 and Diamond 400 reflections and obtained the crystal thicknesses $t_0$ corresponding to focusing conditions $182$ $\mu$m and $80$ $\mu$m, respectively. However, in these cases we obtained even stronger
interference effects. For practical reasons it is better to use crystals and reflections where these interference effects are minimized, due to the fact that oscillations of intensity in the focal region smear the focal point and make application of these crystals and reflections for DFS more problematic. We also considered the cases of Ge 220 and Ge
400 reflections where absorption is much stronger than in Si and Diamond crystals, and interference fringes are not observed. Unfortunately, the intensity of the Laue diffracted beam is also strongly suppressed for Ge crystals at these reflections making Ge crystals inappropriate for DFS applications. Finally, our analysis has shown that for DFS applications the most prominent candidates are Si 220 and Diamond 220 reflections.

\begin{figure}
\centering \includegraphics{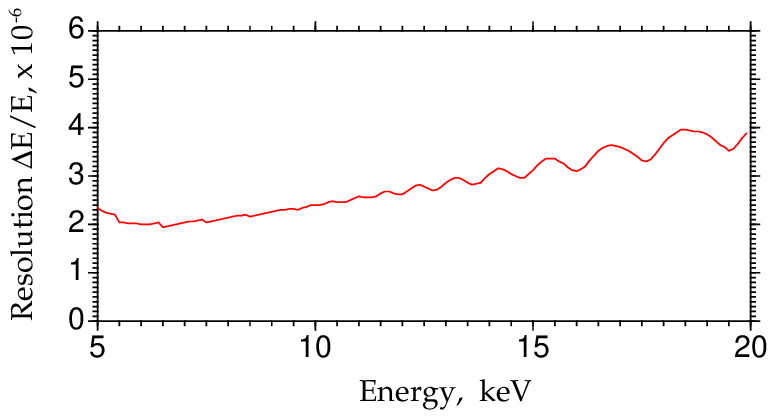}
\caption{Resolution of the DFS as a function of the photon energy for the case of Si 220 reflection. Simulations were performed for the crystal thickness $t=360$ $\mu$m.}
\label{fig6}
\end{figure}

As it follows from our simulations (see Figs. 3(a) and 4(a)) the size of the beam in the focal position is of the order of few microns. The beam profile in the focal position (at crystal thickness $t=t_0$) for two cases of Si and Diamond crystals (220 reflection) is shown in Fig. 5. In this Figure results of computer simulations are presented together with the results of analytical estimates obtained in the previous section. In the case of a Si crystal (Fig. 5, (a)), analytical results obtained by equations (\ref{08}) and (\ref{15}) are slightly different but give an excellent estimate of the beam shape in the focus comparing to the curves obtained from the computer simulations. In the case of a Diamond crystal (see Fig. 5 (b)), both analytical curves completely coincide with each other in the focal region. It is not surprising as both estimates should give good results for a low absorption, that is the case for the Diamond crystal. The approximate curve is close to the one calculated according to the expression (\ref{01}), however, it has slightly larger width. In Table 1 the size of the beam (FWHM) in the focal position for a crystal thickness $t=t_0$, obtained from computer simulations in the case of Si and Diamond crystals, is presented. As it follows from our simulations typical beam sizes are on the order of few microns, that means that the detector should have resolution on the order of a micron, or better.

\begin{center}
\begin{tabular}{|c|c|c|c|c|}
\multicolumn{5}{c}{Table 1. FWHM of the beam in the focal position, ($\mu $m)} \\ \hline
& Si 220 & Si 400 & C 220 & C 400 \\ \hline
Eq. (\ref{01}) & $6.6$ & $7.9$ & $8.2$ & $10.7$ \\ \hline
Eq. (\ref{08}) & $6.7$ & $8.9$ & $9.6$ & $12.7$ \\ \hline
Eq. (\ref{15}) & $7.1$ & $9.3$ & $9.7$ & $12.8$ \\ \hline
\end{tabular}
\end{center}

The energy resolution of the spectrometer for the photon energy of 12.4 keV, reflections 220 and 400 for Si and Diamond crystals was estimated using Eq. (\ref{16}) and is summarized in Table 2. Our results show that at this energy the resolution of DFS is within the range $(2-3)\times 10^{-6}$. As it follows from our simulations, the highest resolution is achieved with the Diamond crystals and 400 reflection. It is interesting to investigate energy resolution as a function of energy at the fixed crystal thickness. Results of our simulations for the energy range of 5 keV to 20 keV, Si crystal, 220 reflection, and crystal thickness $t=360$ $\mu$m are presented in Fig. 6. As can be well seen from this figure, at lower energies about 6 keV resolution is even higher than at the photon energy 12.4 keV. At higher energies resolution decreases with the energy and starts to oscillate. However, up to the energy of 20 keV it stays below the value $5\times 10^{-6}$ that is sufficient for the effective use of DFS. The presence of oscillations in the energy dependence of the resolution can be understood due to the fact that the focusing crystal thickness $t_0$ is energy dependent. As a result, interference between two fields (strongly and weakly absorbing) in the crystal can be observed in the energy range as well.

\begin{center}
\begin{tabular}{|c|c|c|}
\multicolumn{3}{c}{Table 2. The energy resolution of DFS} \\
\multicolumn{3}{c}{at 12.4 keV, ($\times $ $10^{6}$)} \\ \hline
Reflection & Si & C \\ \hline
220 & $3.1$ & $2.5$ \\ \hline
400 & $\quad \;2.7\;\quad $ & $\quad \;2.0\;\quad $ \\ \hline
\end{tabular}
\end{center}

\section{Applications of DFS at XFEL}

Here we will discuss in more detail whether the resolution of the DFS described in the previous sections is sufficient to resolve the fine structure of individual XFEL pulses. To answer this question we simulate the spectral distribution of the incoming XFEL pulse and analyze its shape after Laue diffraction from the crystal. To simulate the initial XFEL energy spectrum we used an approach proposed by Pfeifer et al. (2010) when a double Fourier transform in the spectral and time domains is used starting with the spectrum containing 'white' noise in the phase. The initial XFEL pulse generated according to this procedure is shown in Figure 8(a). We assumed here an incoming photon energy of 12.4 keV, pulse duration $T=100$ fs, and spectral width $\Delta\omega/\omega = 10^{-3}$. As it is well seen from this figure, the generated pulse spectrum  has a typical behavior of a SASE (self-amplified spontaneous emission) XFEL pulse (see, for example, Altarelli et al. (2006)). The spectrum of this pulse obtained after Laue 220 diffraction from a Si crystal is presented in Fig. 8(b). In our simulations we considered the experimental geometry shown in Fig. 2 with the distances $L_{0}=10$ m and $L_{1}=0.1$ m.
It is well seen in Fig. 7 that all main features of the energy spectrum of the incoming XFEL pulse are well resolved and reproduced at the detector. However, due to a finite resolution of DFS, a small additional background can be observed in Fig. 8(b) as well.

In the simulations presented in Fig. 7 a point source and a wide angular divergence of the incident beam was assumed. In principle, the finite source size could reduce the energy resolution of the DFS. As it follows from equation (\ref{02}), in order to keep the resolution value of DFS on the same level, the source size should be significantly smaller then the beam size in the focal position.

\begin{figure}
\centering \includegraphics{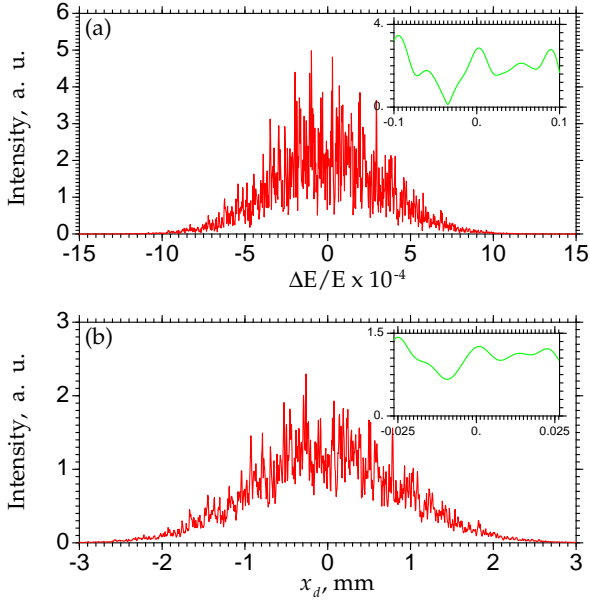}
\caption{(a) Simulated XFEL spectrum with the incoming photon energy of 12.4 keV, pulse duration $T=100$ fs, and spectral width $\Delta\omega/\omega = 10^{-3}$; (b) Corresponding intensity distribution at the detector after 220 diffraction of this pulse from a Si crystal. Insets show an enlarged part of the spectrum. }%
\label{fig7}
\end{figure}
According to our results obtained in the previous section, the source size on the order of, or smaller, than one micron will be sufficient to keep resolution on the level of $10^{-6}$. Since at XFELs the source sizes are on the order of few tenth of microns (Altarelli et al. (2006)), the focusing optics is an important element for an effective use of DFS.

Another important point for an effective use of DFS at XFELs is the possibility to measure the whole energy spectrum of a single pulse. Two conditions should be satisfied in order to reach this goal. The first one puts limitations on the angular divergence $\theta$ of the incoming beam. Taking into account that divergence of the incoming beam should be bigger than the acceptance angle defined by the Bragg condition we obtain $\theta>\tan \theta _{B}(\Delta\omega/\omega )$.
For a typical bandwidth of XFEL sources $\Delta\omega/\omega\sim 10^{-3}$ and reflections considered in our simulations with $\tan \theta_{B}\approx 0.3\div0.7$ we obtain the following condition for a minimum divergence $\theta
_{0}>7\times 10^{-4}$. The angular aperture $\theta _{l}$ of the parabolic absorbing
lens is equal to $\theta _{l}=\delta (\lambda /R\beta )^{1/2}$ (Kohn, 2012), where the complex
index of refraction is assumed as $n=1-\delta +i\beta $, and $R$ is the
curvature radius at the apex of the parabola.
For Si lens and $\lambda =1$ \AA\ absorption and refraction indices are equal to $\delta =3.2\times 10^{-6}$, $%
\beta =3.2\times 10^{-8}$. Therefore from the inequality $\theta _{l}>\theta
$ we arrive to the condition $R<0.8$ $\mu $m that is easy to satisfy in practice.

The second condition follows from the fact that the detector window and the crystal longitudinal size 
have to be larger than the spectrum size $\Delta x_d$. It follows from Eq.(\ref{04}) that $%
\Delta x_{d}>L_{0}\theta _{0}=7$ mm for $L_{0}=10$ m and $\theta
_{0}=7\times 10^{-4}$. For a pixel size 5 $\mu $m the detector has to have
more than 1400 pixels. This can be achieved in modern detectors.

\section{Summary}

In summary, we propose a spectrometer for XFEL based on the effect of diffraction focusing, that theoretically allows to achieve energy resolution about $\Delta E/E\sim 2\times10^{-6}$. This is better than in most existing spectrometers, and should allow, in principle, to resolve the fine structure of individual XFEL pulses. We investigated different crystals and possible reflections and came to the conclusion that the favorable ones are Silicon and Diamond single crystals with the 220 reflection. We obtained analytical solutions for the spectrometer resolution in two limits of a strongly and weakly absorbing crystals and compared them with the computer simulations. In order to measure the full spectrum the incoming beam should have sufficient divergence that can be obtained, for example, by focusing of the incoming beam with CRLs. The proposed spectrometer is easy to operate and can be installed at any hard x-ray beamline at XFEL. By analyzing the resolving power of the DFS we determined that detector resolution should be on the order of one micron, or below what can be achieved by present detectors. Finally, we foresee that spectrometer based on diffraction focusing will be widely used for the spectral analysis of individual pulses of the present and future hard x-ray FELs.

\section{Acknowledgements}

Part of this work was supported by BMBF Project No. 05K10CHG "Coherent Diffraction Imaging and Scattering of Ultrashort Coherent Pulses with Matter" in the framework of the German-Russian collaboration "Development and Use of Accelerator-Based Photon Sources". The work of VGK was partially supported by RFBR grant N.10-02-00047a. We are thankful for a careful reading of the manuscript to H. Franz, A. Bell, U. Lorenz, and A. Singer, as well as to E. Weckert for his interest and support of this project.

\appendix

\section{Geometrical Optics Approach for Description of the Diffraction Focusing}

Here, we will describe a process of diffraction focusing on the basis of the
geometrical optics approximation (see Fig. 8). In geometrical optics the ray
trajectories from the source to the detector are considered. First,
the base trajectory, which is used in the derivation of the equation (\ref{01}) has to be defined.
This trajectory starts from the source and goes to the crystal at the
Bragg angle with the diffracting atomic planes, after taking into account the
refraction at the crystal surface. Inside the crystal it goes along the
atomic planes in the symmetrical case of diffraction. Behind the crystal it goes
along the direction of the reflected beam.
This trajectory is shown in Fig. 8 by the red line. It defines the
origin on the detector coordinate $x_{d}$ axis, i.e. it corresponds to the
point $x_{d}=0$. The integration variable $\eta $ in Eq. (\ref{01}) is
proportional to the angular deviation from this base trajectory. Each value
of $\eta $ corresponds to the ray, and all rays make a total contribution
the detector image. As usual, the trajectory of the ray is defined from the
calculation of the integral (\ref{01}) by means of the stationary phase method.

\begin{figure}
\centering \includegraphics{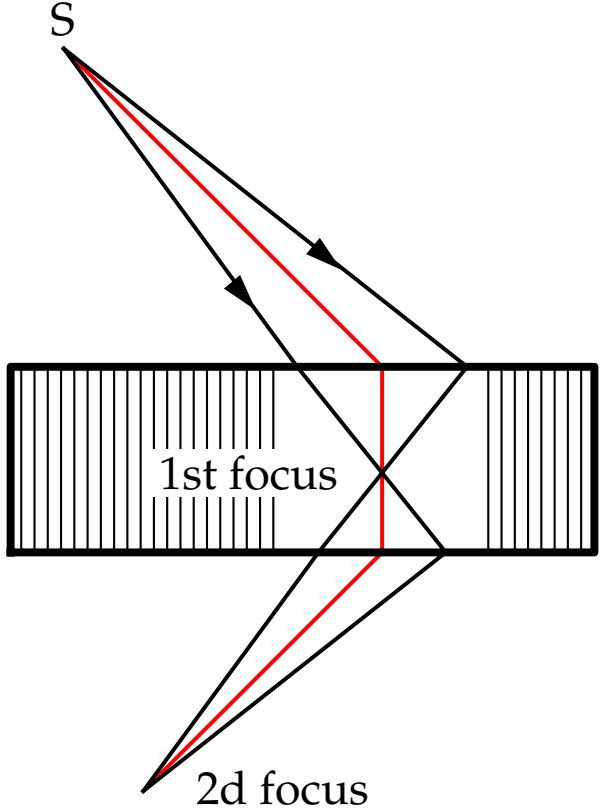}
\caption{Schematic view of diffraction focusing based on geometrical optics approach.}%
\label{fig8}
\end{figure}
The main contribution to the integral corresponds to the region where the
condition%
\begin{equation}
\frac{d\Phi _{\pm }(\eta )}{d\eta }=0  \label{a1}
\end{equation}%
is satisfied. We are interested in the small values of $\eta \ll 1.$
Therefore, we can consider only the linear term in Taylor series over $\eta $%
. In addition, we restrict ourselves by the symmetrical case $\beta =1$.

Let us consider, first, the case in the front of the crystal. Then $L_{1}=t=0$
and from Eqs. (\ref{a1}), (\ref{02}) and (\ref{03}) we have%
\begin{equation}
x_{d}=\frac{|\chi _{rh}|}{S_{B}}L_{0}\eta  .
\label{a2}
\end{equation}%
As it follows from Eq. (\ref{a2}), the coordinate $x_{d}$ is proportional to the angular
deviation parameter $\eta$. It is the usual situation for a spherical wave. Let us consider, in
addition, the crystal plate of definite thickness $t$. Now we obtain two
trajectories according to the two signs in the equation. They belong to two
branches of the dispersion surface which are well known in the theory of
plane wave diffraction in crystals (Authier, 2005).

We are interested in the upper sign, i.e. weakly absorbing wave. For this
wave we obtain the following trajectory equation for small  $\eta \ll 1$%
\begin{equation}
x_{d}=\frac{S_{B}}{2\gamma _{0}}(t_{0}-t)\eta ,\quad t_{0}=\frac{2|\chi
_{rh}|\gamma _{0}}{S_{B}^{\,2}}L_{0}  \label{a3}
\end{equation}%
As it follows from Eq. (\ref{a3}), the coordinate $x_{d}$ decreases with increasing the crystal thickness $t$.
At a certain crystal thickness $t=t_{0}$ the coordinate $x_{d}$  stays zero
for all values of $\eta $ inside the region where $\eta \ll 1$. At this
thickness all rays go to the same point, that means focusing. If the
crystal thickness exceeds that focusing distance $t>t_{0}$, then focusing takes place inside the crystal
(see Fig. 9), and rays become again divergent inside a crystal, but now $x_{d}$ is
negative for a positive $\eta $, and vice versa.

Let us consider now the case $t>t_{0}$ and the distance $L_{1}$ behind the
crystal. We obtain%
\begin{equation}
x_{d}=\left[ \frac{S_{B}}{2\gamma _{0}}(t_{0}-t)+\frac{|\chi _{rh}|}{S_{B}}%
L_{1}\right] \eta  .
\label{a4}
\end{equation}%
Here the first term in the square brackets is negative, but the second term
is positive. Therefore, the second focusing is possible at the definite distance
behind the crystal (see Fig. 8).

\vfill\eject

\end{document}